\documentclass[a4paper,12pt]{article}

\pdfoutput=1
\usepackage[numbers,sort&compress]{natbib}
\usepackage[utf8]{inputenc}
\usepackage[english]{babel}
\usepackage{hyperref}
\hypersetup{
    colorlinks,
    linkcolor={red!50!black},
    citecolor={blue!50!black},
    urlcolor={blue!80!black}
}
\usepackage{amsfonts,amsmath,amssymb,amsthm,mathtools}

\usepackage{graphicx,pgfplots,pgfkeys,pgfmath,bbm,geometry,marginnote,tikz,xcolor,url,enumitem,xspace,listings}
\usetikzlibrary{positioning,arrows,shapes.geometric,calligraphy,decorations,calc}

\newcounter{nthm}[section] 
\theoremstyle{definition}
\newtheorem{definition}[nthm]{Definition}
\newtheorem{remark}[nthm]{Remark}
\theoremstyle{plain}

\newtheorem{lemma}[nthm]{Lemma}
\newtheorem*{lemma*}{Lemma}

\newtheorem{corollary}[nthm]{Corollary}

\lstnewenvironment{algorithm}{
    \lstset{ 
        mathescape=true,
        escapeinside={(*@}{@*)},
        numbers=none, 
        numberstyle=\tiny,
        basicstyle=\footnotesize, 
        keywordstyle=\bfseries\em,
        keywords={,input, output, not, and, or, if, then, else, while, in, do, begin, end,}
        morecomment=[l]{//}, 
        commentstyle=\itshape
    }
}{}

\DeclareMathOperator{\Span}{span}
\newcommand{\bigO}[1]{\ensuremath{\mathcal{O}\left({#1}\right)}}
\newcommand{\Tr}[1]{\ensuremath{\mathrm{Tr} \left( {#1} \right)}}
\newcommand{\abs}[1]{\lvert {#1} \rvert}
\newcommand{\ket}[1]{\lvert {#1} \rangle}
\newcommand{\bra}[1]{\langle {#1} \rvert}
\newcommand{\ketbra}[2]{\ket{#1}\!\bra{#2}}
\newcommand{\braket}[2]{\langle {#1} \!\mid\! {#2} \rangle}

\newcommand{\card}[1]{\lvert {#1} \rvert}

\newcommand{\floor}[1]{\lfloor {#1} \rfloor}

\newcommand{\eye}{\ensuremath{\cdot\!\!(\!\!\!>}}

\newcommand{\cH}{\ensuremath{\mathcal{H}}}

\newcommand{\cP}{\ensuremath{\mathcal{P}}}
\newcommand{\cA}{\ensuremath{\mathcal{A}}}

\newcommand{\cC}{\ensuremath{\mathcal{C}}}
\newcommand{\cW}{\ensuremath{\mathcal{W}_\bot}}

\newcommand{\Nat}{\ensuremath{\mathbb{N}}}
\newcommand{\CTwo}{\ensuremath{\mathbb{C}^2}}
\newcommand{\eg}{\textit{e.g.}\xspace}
\newcommand{\ie}{\textit{i.e.}\xspace}

\title{Weakly measured while loops: \\ \Large peeking at quantum states}

\author{\normalsize Pablo Andr\'es-Mart\'inez\footnote{p.andres-martinez@ed.ac.uk} \ and Chris Heunen\footnote{chris.heunen@ed.ac.uk} \\ \normalsize University of Edinburgh}

\begin{document}

\maketitle

\begin{abstract}
A \emph{while} loop tests a termination condition on every iteration.
On a quantum computer, such measurements perturb the evolution of the algorithm.
We define a while loop primitive using weak measurements, offering a trade-off between the perturbation caused and the amount of information gained per iteration.
This trade-off is adjusted with a parameter set by the programmer.
We provide sufficient conditions that let us determine, with arbitrarily high probability, a worst-case estimate of the number of iterations the loop will run for.
As an example, we solve Grover's search problem using a while loop and prove the quadratic quantum speed-up is maintained.
\end{abstract}

\section{Introduction}

In quantum computer science, \emph{while} loops do not hold centre stage.
They are often used in a trivial way, for instance, when we say that a quantum algorithm is repeated until it succeeds.
On the other hand, there are examples, such as Grover's algorithm, where a quantum operation is applied a fixed number of times, as in a \emph{for} loop.
But, if a subroutine's behaviour is too costly (or even impossible) to predict, we require the use of a while loop to be sure it runs until the termination condition is met.

Measuring a quantum state perturbs it, so there is no way to test a while loop's termination condition without affecting the quantum computation.
In this work, we propose a while loop primitive, which we call \(\kappa\)-while loop, where the detrimental effect of the measurement can be limited.
The key component of a \(\kappa\)-while loop is the use of \emph{weak measurements} to test the termination condition of the loop.
Loosely speaking, tuning the parameter \(\kappa\) -- known as the \emph{measurement strength} -- can reduce the amount of collapse, at the cost of gaining less information from each measurement. 
Under certain conditions, we may use \(\kappa\)-while loops to ``monitor'' a property of the evolving system throughout the iterations, stopping only when success is certain.

We introduce two properties, \emph{active guarantee} and \emph{robustness}, that let us analyze the performance of \(\kappa\)-while loops. As proof of concept, we show that Grover's iteration satisfies these properties, thus allowing us to implement Grover's algorithm using a \(\kappa\)-while loop.
We do not fix the number of iterations in advance, but knowledge of the database's size is still necessary to determine the value of \(\kappa\). 

The structure of the paper is as follows. In Section~\ref{sec:background} we discuss related work and define the notion of \(\kappa\)-measurements. Section~\ref{sec:general} presents the main contribution: the definition of \(\kappa\)-while loops and the properties of \emph{active guarantee} and \emph{robustness}. In Section~\ref{sec:Grover}, we apply our framework to Grover's search problem, providing an algorithm similar to the one previously proposed by Mizel~\cite{Mizel}. We conclude with some discussion in Section~\ref{sec:discussion} and suggest possible applications of \(\kappa\)-while loops beyond quantum search.

\section{Background}
\label{sec:background}

In this section we discuss the two approaches to control flow in quantum programming languages and introduce the concept of \(\kappa\)-measurement and its connection to weak measurements.

\subsection{While loops in quantum computing}
\label{sec:loop_review}

There are multiple examples of quantum programming languages in the literature (see Ref.~\cite{PLSurvey} for a survey).
We can classify these languages into two different paradigms: quantum control flow and classical control flow.
In this section we summarise the most relevant aspects of each paradigm.
For a more in-depth discussion, see Ref.~\cite{YingBook}.

\paragraph{Quantum control flow.} QML~\cite{QML} was the first quantum programming language exhibiting quantum control flow. In this programming language, an if-then-else statement
\begin{algorithm}
                        if$^\circ$ b then $U$ else $W$ end
\end{algorithm}
corresponds to constructing a quantumly controlled gate. If the control qubit \(b\) is \(\ket{1}_b\) then gate \(U\) is applied, and if it is \(\ket{0}_b\) then gate \(W\) is applied. If \(b\) is in a superposition, both branches of the control flow coexist in superposition.

There are multiple problems with QML's original approach~\cite{AlternationPanangaden}, which later works attempted to amend~\cite{ValironFixedPoint,AlternationPanangaden, AlternationYing}.
These works have tackled important issues of the language's semantics, but a persistent problem is the impossibility of implementing unbounded loops (and unbounded recursion).
Each time the control flow forks, a fresh auxiliary qubit is required to act as the control condition, and this qubit gets entangled to the state of the computation.\footnote{In certain situations, the state of the auxiliary qubit can be \emph{uncomputed}, \ie\ returned to its initial value by applying the inverse of part of the computation. However, this is only possible when none of the qubits used to compute the value of \(b\) are affected by \(U\) or \(W\). Thus, auxiliary control qubits cannot be reused in general.}
Therefore, while loops -- where the control flow forks on each iteration -- would have their maximum number of iterations bounded by the size of the quantum memory; a similar problem would arise with unbounded recursion.
Moreover, as pointed out by Linden and Popescu~\cite{HaltingPopescu}, a program with quantum control flow cannot support an unbounded loop where different terms of the superposition exit the loop at different times.
More specifically, if the evolution is required to be unitary, terms that exited the loop at different times cannot interfere with each other, thus the control flow cannot be said to be fully quantum.

\paragraph{Classical control flow.} These quantum programming languages follow the slogan ``quantum data, classical control''~\cite{SelingerPL}. Consider a quantum register \(q\) that holds states from a Hilbert space \(\cH\) and let \(\{M_0,M_1\}\) describe a projective measurement on \(\cH\).
The statement
\begin{algorithm}
                        while $M[q]=0$ do $U$ end
\end{algorithm}
corresponds to applying, on each iteration, unitary \(U\) on the state of register \(q\) followed by the projective measurement \(\{M_0,M_1\}\)~\cite{YingLoop}.
If the outcome of the measurement is \(0\), the loop keeps iterating; otherwise it halts and the next statement of the program is executed.
There is no superposition of commands, as the loop either halts or continues iterating, hence the term classical control flow.
This paradigm is simple to realise on physical devices -- it is just a quantum chip controlled by a classical computer -- and its semantics are better understood.
The drawback of this approach is that applying a measurement on each iteration perturbs the state being observed, and thus alters the computation itself.

In this work we propose weakly measured while loops, an example of classical control flow where the effect of the collapse is kept below a threshold.
Under certain conditions (see Section~\ref{sec:main}), we prove that the performance of the quantum evolution is unaffected.

\subsection{Weak measurements}
\label{sec:weak_meas}

Roughly speaking, a weak measurement is ``a measurement which gives very little information about the system on average, but also disturbs the state very little''~\cite{ToddTutorial}.
The field of quantum feedback control uses weak measurements (often, continuous measurements) to monitor a state. The stream of measurement outcomes is used to control the strength of a Hamiltonian that corrects the system; see Ref.~\cite{QuantumFeedbackSurvey} for a survey.
Our approach is inspired on these ideas, contextualised for their application to algorithm design.
We restrict ourselves to the discrete-time setting and define a particular kind of parametrised measurement, the \(\kappa\)-measurement, that behaves as a weak measurement when \(\kappa\) is small.

Let \(\cH\) be a Hilbert space; we wish to apply a measurement to test whether a state \(\psi \in \cH\) satisfies certain property.
Let \(B\) be an orthonormal basis of \(\cH\) such that the outcome of the measurement on each \(b \in B\) is deterministic; we can characterise the property we wish to measure via a function \(Q \colon B \to \{0,1\}\) which determines whether \(b \in B\) satisfies it \(Q(b) = 1\) or not \(Q(b) = 0\).
We refer to \(Q\) as the \emph{predicate} to be tested by the measurent: in computer science, a predicate is a function assigning to each element of a set a truth value.
Assume the existence of a unitary \(O_Q \colon \cH \otimes \CTwo \to \cH \otimes \CTwo\) acting as the \emph{oracle} of predicate \(Q\):
\begin{equation}
\label{eq:oracle}
  O_Q \ket{x,p} = \ket{x,p \oplus Q(x)}.
\end{equation}
Fix a value of parameter \(\kappa \in [0,1]\) and let \(\cP = \Span\{\bot,\top\}\) be an auxiliary space known as the \emph{probe}. We define a unitary \(E_{\kappa,Q}\) that rotates the state of the probe an amount according to \(\kappa\) only if the state in \(\cH\) satisfies \(Q\):
\begin{align} \label{eq:E_kQ}
  E_{\kappa,Q} &= (O_Q^\dagger \otimes I_\cP) \, (I_\cH \otimes \Lambda(R_\kappa)) \, (O_Q \otimes I_\cP) \\
  \Lambda(R_\kappa) &= \ketbra{0}{0} \otimes I_\cP + \ketbra{1}{1} \otimes R_\kappa \\
  R_\kappa &= \begin{pmatrix}
               \sqrt{1 - \kappa} &  \sqrt{\kappa} \\
               \sqrt{\kappa}     & -\sqrt{1 - \kappa} \\
              \end{pmatrix},
\end{align}
Notice that the auxiliary qubit where the oracle \(O_Q\) writes the result is restored to its initial state by \(O_Q^\dagger\), so it may be reused. We initialise it to \(\ket{0}\) and omit it in further discussions.

\begin{remark}
  In~\eqref{eq:E_kQ} the only purpose of \(R_\kappa\) is to send \(\ket{\bot}\) to \(\alpha \ket{\bot} + \beta \ket{\top}\) so that \(\lvert \beta \rvert^2 = \kappa\). The definition of \(R_\kappa\) provided is just one of infinitely many possible choices. In general, \(Z(\theta) R_\kappa Z(\theta')\) for any \(\theta\) and \(\theta'\) is a valid choice for this unitary, where \(Z(\theta)\) is a \(Z\)-rotation of angle \(\theta\).
\end{remark}

\begin{definition}
A \(\kappa\)-measurement of predicate \(Q\) may be applied on any density matrix \(\rho\) by the following procedure:
\begin{itemize}
  \item apply the unitary \(E_{\kappa,Q} \colon \cH\otimes\cP \to \cH\otimes\cP\) on the state \(\rho \otimes \ketbra{\bot}{\bot}\),
  \item apply a measurement on the probe space \(\cP\) determined by projectors
  \begin{equation} \label{eq:M_P}
    M_\cP = \{I_{\cH}\otimes\ketbra{\bot}{\bot},I_{\cH}\otimes\ketbra{\top}{\top}\}.
  \end{equation}
\end{itemize}
\end{definition}

In essence, the state of the probe is entangled with the result of the oracle so that, when the probe is measured, we may learn something about the state in \(\cH\).
The degree of the entanglement is determined by the parameter \(\kappa\): for \(\kappa = 1\) the entanglement is maximal, whereas for \(\kappa = 0\) there is no entanglement at all.

\begin{remark} \label{rmk:known_probe_outcome}
After applying a measurement, we may describe the resulting state as a mixed state on the space \(\cH \otimes \cP\). However, in doing so we would be omitting the information we have obtained from the classical outcome of the measurement: we in fact know what the state in \(\cP\) is.
For our purposes, it is more elucidating to describe the two possible outcomes separately, providing both the resulting state in \(\cH\) when the measurement readout is \(\top\) and that when the readout is \(\bot\).
Notice that, if the state in \(\cH\) prior to measurement were a pure state, both the outcome after we read \(\bot\) or \(\top\) will be pure states; this fact will be used in Section~\ref{sec:Grover}.
\end{remark}

If the outcome of the \(\kappa\)-measurement is \(\top\), we are certain that the state left in space \(\cH\) satisfies predicate \(Q\). On the other hand, outcome \(\bot\) provides no definitive information about \(Q\) (except when \(\kappa = 1\)). For any state \(\rho\), the probability of outcome \(\top\) is
\begin{align}
\label{eq:pTop}
  p_\top(\rho) &= \kappa \cdot p_Q(\rho) \\
\label{eq:pQ}
  p_Q(\rho) &= \Tr { O_Q (\rho \otimes \ketbra{0}{0}) O_Q^\dagger \ (I_\cH \otimes \ketbra{1}{1}) }
\end{align}
where \(p_Q(\rho)\) is the probability of predicate \(Q\) being satisfied by \(\rho\). The smaller \(\kappa\) is, less likely it is that we read outcome \(\top\). In exchange, the map corresponding to output \(\bot\):
\begin{align}
\label{eq:action_W}
  \cW(\rho) &= \frac{W_\bot (\rho \otimes \ketbra{\bot}{\bot}) W_\bot^\dagger}{1 - p_\top(\rho)} \\
  W_\bot &= (I_{\cH} \otimes \ketbra{\bot}{\bot})\,E_{\kappa,Q}
\end{align}
does \emph{not} fully collapse the state to the subspace where \(Q\) is unsatisfiable.


\section{A weakly measured while loop}
\label{sec:general}

This section contains the main contribution of our paper: we define \(\kappa\)-while loops and introduce the properties of \(\cA\)-guarantee and robustness. In short, a \(\kappa\)-while loop is a classically controlled while loop where the test of the termination condition is realised by a \(\kappa\)-measurement.

\subsection{Motivation}
\label{sec:motivation}

Fix a predicate \(Q\), and let \(\cC\) be a completely positive (CP) map acting on \(\cH\); we will refer to \(\cC\) as the \emph{body} of the loop and write \(D(\cH)\) for the set of density operators on \(\cH\).

\begin{definition}
Let \(\sigma \in D(\cH)\) be an arbitrary state. Define a function \(\cA_{Q,\sigma} \colon \Nat \to \{0,1\}\) by
\begin{equation}
  \cA_{Q,\sigma}(n) \iff p_Q(\cC^n(\sigma)) > \frac{1}{2}
\end{equation}
where \(p_Q\) is given in~\eqref{eq:pQ}, and \(\cC^n(\sigma) = \overbrace{\cC(\cdots (\cC}^{\footnotesize n\text{ times}}(\sigma)))\).
If \(\cA_{Q,\sigma}(n)\) is satisfied, we say \(n\) is an \(\cA\)-iteration (read as \emph{active iteration}).
\end{definition}

If we somehow identified an \(\cA\)-iteration \(m\), we may use a simple approach to find a state satisfying \(Q\): apply \(\cC^m(\sigma)\) and then perform a projective measurement; if the outcome does not satisfy \(Q\), repeat the process from the initial state \(\sigma\). Thanks to \(\cA_{Q,\sigma}(m)\) being satisfied, the probability of succeeding at some point within \(k\) restarts is \(1 - 1/2^k\) so the probability of success quickly approaches \(1\) as \(k\) increases.
This is an efficient approach whenever a small \(\cA\)-iteration \(m\) is known. 
In fact, this is how the standard Grover's algorithm works, choosing an iteration \(m\) where \(p_Q\) is maximised and extremely close to \(1\).

However, this measure-restart strategy can only be implemented if we already know when \(\cA\)-iterations will occur.
The distribution of \(\cA\)-iterations may become unpredictable as soon as some randomness is introduced in the body of the loop.
For instance, consider the standard Grover's algorithm being implemented in a faulty machine where, without notice, the quantum memory may collapse to its initial state. 
If such collapse happens mid-computation, the evolution is effectively restarted, but the algorithm -- oblivious to the collapse -- continues for only the remaining fixed number of iterations.
In contrast, a version of Grover's algorithm that uses a \(\kappa\)-while loop (such as the one we describe in Section~\ref{sec:Grover}) will, by definition, keep iterating until it succeeds to find the target state.
Although a contrived example, this illustrates how a while loop could provide reliability against unpredictable behaviour.
More realistic situations that would lead to an unpredictable distribution of \(\cA\)-iterations could come from algorithms that, by design, apply mid-computation measurements or require interaction with the environment.
Other situations where the use of \(\kappa\)-while loops may be advantageous are discussed in Section~\ref{sec:beyond_quantum_search}.

In this section we present the concept of \(\kappa\)-while loops as an abstract quantum programming construct.
In Section~\ref{sec:Grover} we present a version of Grover's algorithm that uses a \(\kappa\)-while loop and maintains the quadratic quantum speed-up.
In the future, we hope to apply \(\kappa\)-while loops to practical problems where the distribution of \(\cA\)-iterations cannot be predicted.
To this end, we provide sufficient conditions that let us determine, for arbitrarily high probability, a worst-case estimate of the number of iterations the loop will run for.
Remarkably, these conditions do not require us to know the precise distribution of \(\cA\)-iterations, but only a guarantee of their proportion throughout the algorithm in the worst-case scenario.

\subsection{The \(\kappa\)-while loop}
\label{sec:main}

In Figure~\ref{fig:kappa_while} we propose the syntax for the \(\kappa\)-while loop and its implementation on a quantum programming language with classical control flow. The syntax of the \(\kappa\)-while loop is meant to be read as ``repeat \(\cC\) while it is \emph{not certain} that \(Q\) is satisfied''.
On each iteration, the value of the state \(\rho\) will be updated to \(\cC(\rho)\), followed by a \(\kappa\)-measurement of predicate \(Q\) (see Section~\ref{sec:weak_meas}). If the outcome of the \(\kappa\)-measurement is \(\bot\), the loop keeps iterating; the state \(\rho\) becomes \(\cW(\rho)\), collapsing some of the quantum information. Otherwise, outcome \(\top\) halts the loop and we succeed in obtaining a state \(\rho\) that satisfies \(Q\).

\begin{figure}
\begin{tikzpicture}
  \node (code_alias) {
    \begin{algorithm}
        $\boldsymbol{\kappa}-$while $Q[\rho] \not = 1$ do
          $\rho$ $\gets$ $\cC(\rho)$
        end
    \end{algorithm}
  };
  \node[right=5mm of code_alias] (code_def) {
    \begin{algorithm}
        $q$ $\gets$ $\ketbra{\bot}{\bot}$
        while $M_\cP[q] = \bot$ do
          $\rho$ $\gets$ $\cC(\rho)$
          $\rho \otimes q$ $\gets$ $E_{\kappa,Q}(\rho \otimes q)$
        end
    \end{algorithm}
  };
\end{tikzpicture}
\caption{\emph{Left:} the syntax we use to represent a \(\kappa\)-while loop; \(\kappa \in [0,1]\) is a parameter set by the programmer and \(Q\) is the predicate to be measured. \emph{Right:} the pseudocode that implements the \(\kappa\)-while loop on a programming language with classical control flow; \(E_{\kappa,Q}\) and \(M_\cP\) are defined in Section~\ref{sec:weak_meas}.}
\label{fig:kappa_while}
\end{figure}

\begin{definition}
\label{def:guarantee}
A function \(f \colon \Nat \to \Nat\) is a \emph{guarantee} of \(\cA_{Q,\sigma}\) if:
\begin{equation}
\label{eq:guarantee}
  \forall n \in \Nat, \;\; \exists m \leq f(n)\colon \quad n = \abs{\{ k \in \Nat \mid k < m,\, \cA_{Q,\sigma}(k) \}}
\end{equation}
\end{definition}

If such a function \(f\) exists, we are promised that there will be at least \(n\) active iterations within the first \(f(n)\) applications of \(\cC\). In principle, this need not be a tight bound; for instance, we may know that, within the first thousand iterations, there will be at least ten \(\cA\)-iterations. In that case, a valid guarantee of \(\cA_{Q,\sigma}\) would be \(f(1)=f(2)=\dots=f(10) = 1000\). This gives us little information about when any of these \(\cA\)-iterations actually occur.

\begin{definition}
\label{def:robustness}
The evolution induced by a CP map \(\cC\) on a state \(\sigma\) is said to be \(\varepsilon\)-\emph{robust} to \(\kappa\)-measurements of predicate \(Q\) if there is a function \(g \colon \Nat \to \Nat\) such that:
\begin{equation}
  \forall n \in \Nat, \;\; \exists m \leq g(n) \colon \quad \abs{p_Q(\cC^n(\sigma)) - p_Q((\cW\cC)^m(\sigma))} \leq \varepsilon
\end{equation}
for \(\varepsilon \ll 1/2\). The function \(g\) is called a \emph{witness} of the robustness.
\end{definition}

\begin{lemma}
\label{lem:comp_guarantee}
If \(f\) is a guarantee of \(\cA_{Q,\sigma}\) and \(g\) witnesses that \(\cC\) is \(\varepsilon\)-robust, then \(g \circ f\) is a guarantee of:
\begin{equation}
  \cA'_{Q,\sigma}(n) \iff p_Q((\cW\cC)^n(\sigma)) > \frac{1}{2} - \varepsilon
\end{equation}
\end{lemma} 
\begin{proof}
After the first \(f(n)\) applications of \(\cC\), there will be at least \(n\) \(\cA\)-iterations. If we switch to weakly measured iterations \(\cW\cC\), robustness tells us that within the first \(g(f(n))\) iterations of the loop we will find those \(n\) active iterations again, but the probability \(p_Q\) might differ by \(\varepsilon\), so it is at least \(p_Q > 1/2 - \varepsilon\). Thus, for any \(n \in \Nat\), there is $m \leq g(f(n))$ satisfying:
\begin{equation}
n = \abs{\{ k \in \Nat \mid k < m, \cA'_{Q,\sigma}(k) \}}
\end{equation}
This concludes the proof.
\end{proof}

Lemma~\ref{lem:comp_guarantee} guarantees that at least \(n\) of the first \(g(f(n))\) iterations of the \(\kappa\)-while loop will be \(\cA'\)-iterations. For each of these \(\cA'\)-iterations, the probability of outcome \(\top\) is at least \(\kappa(1/2 - \varepsilon)\). Within \(N\) \(\cA'\)-iterations the probability of loop termination is:
\begin{equation}
P_{succ} \geq 1 - \big(1 - \kappa(1/2 - \varepsilon)\big)^N
\end{equation}
If \(x \in (0,1)\), then \((1 - x)^\frac{1}{x} < \tfrac{1}{e}\). Therefore, if the algorithm is allowed to run for at least \(N = \frac{2}{\kappa(1-2\varepsilon)}\) \(\cA'\)-iterations, the probability of success is:
\begin{equation}
  P_{succ} \geq 1 - \tfrac{1}{e} > \tfrac{1}{2}
\end{equation}
Therefore, with probability higher than \(1/2\), our \(\kappa\)-while loop will halt within its first 
\begin{equation}
  T = g\left(f\left(\frac{2}{\kappa(1-2\varepsilon)}\right)\right)
\end{equation}
iterations.
More generally, for any \(c \in \Nat\), the probability of success after \(c N\) \(\cA'\)-iterations is greater than \(1 - 1/e^c\). Therefore, with arbitrarily high probability, the loop halts successfully within \(T_c = g(f(cN))\) iterations for small \(c\).

\begin{remark}
Definition~\ref{def:guarantee} can be weakened so that instead of \(f\) being a `deterministic' guarantee of \(\cA_{Q,\sigma}\), we only impose that \(f(n)\) satisfies~\eqref{eq:guarantee} with probability higher than \(\eta\).
In such a case, after \(T_c\) iterations we can achieve a success probability arbitrarily close to \(\eta\).
\end{remark}

\section{An example: Grover's algorithm}
\label{sec:Grover}

In Grover's search problem~\cite{Grover}, we are given an unsorted set of elements \(B\), about which we know no structure or heuristics, and a predicate \(\chi \colon B \to \{0,1\}\) that satisfies \(\chi(\star) = 1\) for only one element \(\star \in B\).
We are tasked with finding this marked element \(\star\).
The standard Grover's algorithm defines an iteration operator \(G\) (using an oracle of \(\chi\)) and applies it a fixed number of times \(K\) on an initial state \(\ket{\psi}\).
Afterwards, a PVM on the basis \(B\) is applied, finding the marked element with high probability. 
In this section we discuss a different approach to Grover's search problem where a \(\kappa\)-while loop is used instead.

\begin{figure}
\begin{tikzpicture}
\node (walk) {
  \begin{tikzpicture}
    \node[trapezium,draw=black,thick,trapezium angle=45,minimum width=15mm,shape border rotate=90] (BS) {\scriptsize \(U_\cP\)};
    \coordinate[below=4mm of BS.east] (BS-0); \node[left=-0.5mm of BS-0] {\tiny \(\bot\)};
    \coordinate[above=4mm of BS.east] (BS-1); \node[left=-0.5mm of BS-1] {\tiny \(\top\)};
    \node[trapezium,draw=black,thick,trapezium angle=45,minimum width=15mm,shape border rotate=270,right=30mm of BS] (BSd) {\scriptsize \(U_\cP^{\dagger}\)};
    \coordinate[below=4mm of BSd.west] (BSd-0); \node[right=-0.5mm of BSd-0] {\tiny \(\bot\)};
    \coordinate[above=4mm of BSd.west] (BSd-1); \node[right=-0.5mm of BSd-1] {\tiny \(\top\)};
    \node[right=7mm of BS-1] (meas) {\large \(\eye\)};
    \coordinate[left=7mm of BSd-1] (0);
    \coordinate[above=1.6mm of 0] (0up);
    \coordinate[below=1.6mm of 0] (0down);
    \node[rectangle,draw=black,thick,right=12mm of BS-0] (G) {\small \(G\)};
    \node[rectangle,draw=black,thick,right=9mm of BSd] (E) {\small \(E_{\kappa,\chi}\)};
    \coordinate[right=4mm of E] (rightE);
    \coordinate[below=15mm of rightE] (belowE);
    \coordinate[left=4mm of BS] (leftBS);
    \coordinate[below=15mm of leftBS] (belowBS);
    \draw (0up) edge[ultra thick] (0down);
    \draw (0) --node[midway,pin={[pin distance=-2mm, pin edge={opacity=0}] above : \scriptsize\(\cH\)}] {\scriptsize \(>\)} (BSd-1);
    \draw (BS-1) --node[midway,pin={[pin distance=-2mm, pin edge={opacity=0}] above : \scriptsize\(\cH\)}] {\scriptsize \(>\)} (meas);
    \draw (BS-0) --node[midway,pin={[pin distance=-2mm, pin edge={opacity=0}] below : \scriptsize\(\cH\)}] {\scriptsize \(>\)} (G.west);  
    \draw (G.east) --node[midway,pin={[pin distance=-2mm, pin edge={opacity=0}] below : \scriptsize\(\cH\)}] {\scriptsize \(>\)} (BSd-0); 
    \draw (BSd.east) --node[midway] {\scriptsize \(>\)} (E.west); 
    \draw (E.east) --node[right] {\scriptsize \(>\)} (rightE); 
    \draw (rightE) edge[out=0,in=0,looseness=1.5] (belowE);
    \draw (belowE) --node[midway,pin={[pin distance=-2mm, pin edge={opacity=0}] below : \scriptsize\(\cH \otimes \cP\)}] {\scriptsize \(<\)} (belowBS);
    \draw (belowBS) edge[out=180,in=180,looseness=1.5] (leftBS);
    \draw (leftBS) --node[midway] {\scriptsize \(>\)} (BS.west);
  \end{tikzpicture}
};
\node[left=5mm of walk] (code) {
  \begin{algorithm}
  input: $\chi$, $\ket{\psi}$
  output: $q$
  begin
    $q$ $\gets$ $\ket{\psi}$
    $\boldsymbol{\kappa}-$while $\chi[q] \not = 1$ do
      $q$ $\gets$ $G(q)$
    end
  end
  \end{algorithm}
};
\end{tikzpicture}
\caption{\emph{Left:} pseudocode describing our algorithm. The while loop is controlled by a \(\kappa\)-measurement as described in Section~\ref{sec:main}. \emph{Right:} the algorithm's information flow; \(E_{\kappa,\chi}\) is given in Section~\ref{sec:weak_meas}, \(U_\cP\) applies the canonical isomorphism \(\cH \otimes \cP \cong \cH \oplus \cH\), separating with respect to the orthogonal basis \(\{\ket{\bot},\ket{\top}\}\) of \(\cP\). 
}
\label{fig:code_walk}
\end{figure}
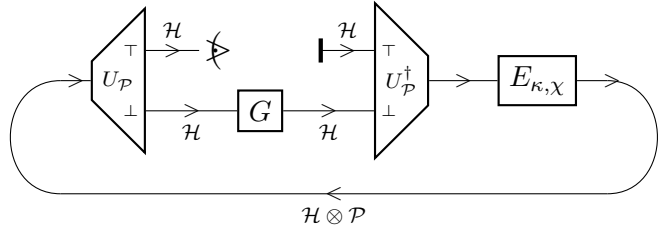

The algorithm's pseudocode is given in Figure~\ref{fig:code_walk}. When the \(\kappa\)-while loop halts, we are \emph{certain} that the state is \(\ket{\star}\).
We do not fix the number of times \(G\) is applied; instead, we fix the measurement strength: \(\kappa\).
We will see that for \(\kappa \approx \card{B}^{-1/2}\) the loop terminates within \(cK\) iterations with arbitrarily high probability, where \(c\) is a small constant and \(K\) is the number of times \(G\) is applied in the standard algorithm.
Thus, the quadratic quantum speed-up of Grover's algorithm is preserved.

It is important to remark that, in the case of Grover's iterator \(G\), we can easily predict when the active iterations occur (see Section~\ref{sec:motivation}).
Therefore, \(\kappa\)-while loops do not provide an algorithmic advantage on this problem.
Instead, we present Grover's problem as a simple proof of concept of how while loops may be used in quantum algorithms without destroying the quantum speed-up. 
This approach to Grover's problem was first proposed by Mizel~\cite{Mizel}.
This section reformulates Mizel's results in the broader framework of Section~\ref{sec:main}.

\subsection{Standard Grover's algorithm}
\label{sec:std_Grover}

The standard Grover's algorithm acts on a quantum state in \(\cH = \Span{B}\), starting from the uniform superposition \(\ket{\psi}\). Let \(\ket{\psi_1} = \ket{\star}\) be the target state and 
\begin{equation}
  \ket{\psi_0} = \tfrac{1}{\sqrt{\card{B}-1}} \sum_{b \in B-\{\star\}} \ket{b}.
\end{equation}
For any \(a \in [0,2\pi)\), define the state \(\ket{a}\) as
\begin{equation} 
\label{eq:angle_state}
  \ket{a} = \cos{a}\,\ket{\psi_0} + \sin{a}\,\ket{\psi_1}.
\end{equation} 
Let \(\alpha = \arcsin{\abs{\braket{\star}{\psi}}} \approx \card{B}^{-1/2}\). Notice that \(\ket{\alpha} = \ket{\psi}\).

For any state \(\ket{\varphi} \in \cH\) we refer to its reflection operator as \(S_\varphi\), given by:
\begin{equation}
  S_\varphi = 2\ketbra{\varphi}{\varphi} - I_\cH
\end{equation}
Grover's iteration is given by the operator
\begin{equation}
\label{eq:Qchi}
  G = S_\psi \, S_{\psi_0}
\end{equation}
where \(S_{\psi_0}\) is implemented using a single call to the oracle of \(\chi\).

On a state \(\ket{a}\), the action of Grover's iteration \(G\) is:
\begin{equation}
\label{eq:action_G}
  G\ket{a} = S_\psi \, S_{\psi_0}\ket{a} = S_\psi\ket{-a} = \ket{a+2\alpha}
\end{equation}
In general, after \(k\) iterations:
\begin{equation}
  G^k\ket{\psi} = \cos{(\alpha + 2k\alpha)}\ket{\psi_0} + \sin{(\alpha + 2k\alpha)}\ket{\psi_1}
\end{equation}
The standard Grover's algorithm applies \(G\) a total of \(K = \floor{(\pi\sqrt{\card{B}})/4}\) times on \(\ket{\psi}\) so that the amplitude of \(\ket{\psi_1}\) is maximised. Then, the state is measured on the basis \(B\), finding the marked element \(\star\) with high probability.

\subsection{While loop approach}
\label{sec:weak_Grover}

In this section we discuss the algorithm given in Figure~\ref{fig:code_walk}. To show that it retains the quantum speed-up, we provide a guarantee of \(\cA_\chi\) and prove that \(G\) is robust to \(\kappa\)-measurements of \(\chi\), as described in Section~\ref{sec:main}.
Our framework (Section~\ref{sec:main}) is general enough to deal with mixed states and CP maps but, in this case, the body of the \(\kappa\)-while loop -- the Grover operator \(G\) -- is unitary and both the initial state \(\ket{\psi}\) and the target \(\ket{\star}\) are pure states.
Therefore, as discussed in Remark~\ref{rmk:known_probe_outcome}, all states involved in our analysis will be pure states.

\begin{lemma} \label{lem:G_guarantee}
The function \(f \colon \Nat \to \Nat\) given by
\begin{align}
  f(n) &= 2n+K \\
  K &= \floor{\tfrac{\pi\sqrt{\card{B}}}{4}}
\end{align}
is a guarantee of \(\cA_\chi\).
\end{lemma} \begin{proof}

By definition~\eqref{eq:angle_state} of \(\ket{a}\), 
\begin{equation}
  p_\chi(\ket{a}) = \abs{\braket{\star}{a}}^2 = \sin^2 a.
\end{equation} 
For two quarters of the circle, in particular whenever \(a \in (\tfrac{\pi}{4},\tfrac{3\pi}{4})\) or \(a \in (\tfrac{5\pi}{4},\tfrac{7\pi}{4})\), we have \(p_\chi(\ket{a}) > \tfrac{1}{2}\).
Moreover, \(G\ket{a} = \ket{a+2\alpha}\) according to~\eqref{eq:action_G}, so the angle is increased a constant amount on each iteration.
Thus, within \(m \in \Nat\) iterations, approximately half of them are active. More precisely, we know that the number of \(\cA\)-iterations is 
\begin{equation}
  n = \frac{m \pm K}{2}
\end{equation}
where the error margin \(K\) refers to the number of applications of \(G\) it takes to traverse a quarter of the circle, which is the longest interval of consecutive iterations not satisfying \(\cA_\chi\).
Solving for \(m\) and taking the worst case scenario, we find that within \(m = 2n + K\) iterations it is guaranteed that \(n\) are active.
\end{proof}

\begin{figure}
\centering
\begin{tikzpicture}
  \node (fstQuad) {
    \begin{tikzpicture}
    \draw[-angle 60,black!65] (0,0) -- (5,0) node[midway,black,xshift=3pt,yshift=-10pt] {\tiny \(\cos{a} \ \ket{\psi_0}\)};
    \draw[-angle 60,black!65] (5,0) -- (5,3.5) node[midway,black,rotate=90,yshift=-11pt] {\tiny \(\sin{a} \ \ket{\psi_1}\)};
    \draw[-angle 60,black!65] (5,0) -- (5,2.5) node[midway,black,rotate=90,xshift=-3pt,yshift=9pt] {\tiny \(\xi \sin{a} \ \ket{\psi_1}\)};
    \draw[-latex,semithick] (0,0) -- (5,3.5) node[midway,yshift=20pt,xshift=10pt] {\scriptsize \(\ket{a}\)};
    \draw[-latex,semithick] (0,0) -- (5,2.5) node[midway,yshift=2pt,xshift=30pt] {\scriptsize \(W_\bot\ket{a}\)};
    \draw[densely dashed] (1,0) arc (0:35:1) node[midway,xshift=7pt] {\tiny \(a\)};
    \draw[densely dashed] (27:2) arc (27:35:2) node[midway,xshift=6pt,yshift=3pt] {\tiny \(\theta\)};
    \end{tikzpicture}
  };
  \node[right=15mm of fstQuad] (sndQuad) {
    \begin{tikzpicture}
    \draw[-angle 60,black!65] (0,0) -- (-5,0) node[midway,black,xshift=-25pt,yshift=-9pt] {\tiny \(\cos{a} \ \ket{\psi_0}\)};
    \draw[-angle 60,black!65] (-5,0) -- (-5,3.5) node[midway,black,rotate=90,xshift=-20pt,yshift=9pt] {\tiny \(\sin{a} \ \ket{\psi_1}\)};
    \draw[-angle 60,black!65] (-5,0) -- (-5,2.5) node[midway,black,rotate=90,xshift=-27pt,yshift=-7pt] {\tiny \(\xi \sin{a} \ \ket{\psi_1}\)};
    \draw[-latex,semithick] (0,0) -- (-5,3.5) node[midway,yshift=18pt,xshift=-20pt] {\scriptsize \(\ket{a}\)};
    \draw[-latex,semithick] (0,0) -- (-5,2.5) node[midway,yshift=2pt,xshift=-45pt] {\scriptsize \(W_\bot\ket{a}\)};
    \draw[dashdotted,black!65] (0,0) -- (1,0);
    \draw[densely dashed] (0.5,0) arc (0:145:0.5) node[midway,yshift=2pt] {\tiny \(a\)};
    \draw[densely dashed] (145:2) arc (145:153:2) node[midway,xshift=-13pt,yshift=3pt] {\tiny \(\theta\)};
    \end{tikzpicture}
  };
  \node[above left=-2mm and -5mm of fstQuad] (a) {\small \emph{i)}};
  \node[above left=-2mm and -2mm of sndQuad] (b) {\small \emph{ii)}};
\end{tikzpicture}
\caption{Geometric relation between the angle \(a\) before weak measurement and the offset \(\theta(a,\kappa)\) after measuring outcome \(\bot\). The construction is provided when \emph{i)} \(a\) is in the first quadrant and when \emph{ii)} \(a\) is in the second quadrant. 
}
\label{fig:theta_triangle}
\end{figure}
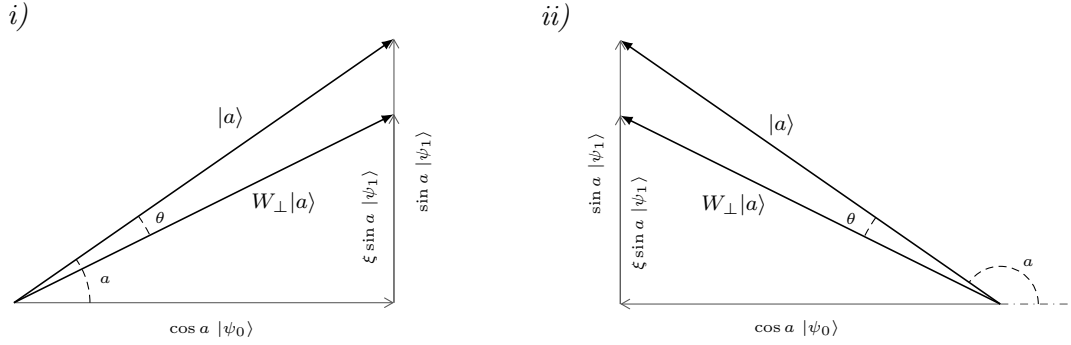

Consider applying a \(\kappa\)-measurement at iteration \(n\): if the outcome is \(\top\), the state becomes \(\ket{\star}\), otherwise it suffers a collapse towards \(\ket{\psi_0}\) determined by the projector \(W_\bot\) from Section~\ref{sec:weak_meas}.
\begin{equation} \label{eq:collapse}
\begin{aligned}
  W_\bot \ket{a,\bot} &=  \cos{a}\,W_\bot\ket{\psi_0,\bot} + \sin{a}\,W_\bot\ket{\psi_1,\bot} \\
   &= \cos{a}\,\ket{\psi_0,\bot} + \sin{a}\,(I_\cH \otimes \ketbra{\bot}{\bot})\,E_{\kappa,\chi}\ket{\psi_1,\bot} \\
   &= \cos{a}\,\ket{\psi_0,\bot} + \sin{a}\,\sqrt{1 - \kappa}\,\ket{\psi_1,\bot}
\end{aligned}
\end{equation}
Thus, the output is still in a superposition between \(\ket{\psi_0}\) and \(\ket{\psi_1}\).
For any \(\ket{a}\) we may describe the action of \(W_\bot\) on it as \(W_\bot\ket{a} \propto \ket{a - \theta(a,\kappa)}\).
Notice that \(W_\bot\) returns an unnormalised state; the actual CPTP map \(\cW\) from~\eqref{eq:action_W} takes care of the normalization, so:
\begin{equation}
  \ket{a} \xmapsto{\cW} \ket{a - \theta(a,\kappa)}
\end{equation}
Figure~\ref{fig:theta_triangle} shows a geometric construction of \(\theta(a,\kappa)\) according to~\eqref{eq:collapse}.
We now find an upper bound of \(\theta(a,\kappa)\); this will be used to prove robustness of \(G\).
To reduce clutter, we use the shorthand:
\begin{equation}\label{eq:xi}
  \xi = \sqrt{1 - \kappa}
\end{equation}

\begin{lemma} \label{lem:theta_bound}
For any \(\kappa\) and angle \(a\)
\begin{equation}
  \abs{\theta(a,\kappa)} \leq \arcsin \left( \frac{1-\xi}{1+\xi} \right).
\end{equation}
\end{lemma} \begin{proof}
See Appendix~\ref{app:bound_proof}.
\end{proof}

\begin{corollary} \label{cor:theta_leq_kappa}
For any angle \(a\), \(\abs{\theta(a,\kappa)} \leq \arcsin \kappa\).
\end{corollary}\begin{proof}

By definition, \(\kappa \in [0,1]\). Using this fact, together with~\eqref{eq:xi} and simple algebra, we can prove that
\begin{equation} \label{eq:aux_leq_kappa}
  0 \leq \frac{1-\xi}{1+\xi} \leq \kappa.
\end{equation}
Then, the corollary follows from Lemma~\ref{lem:theta_bound}.
\end{proof}

In a \(\kappa\)-measurement of \(\kappa \approx 1\) the collapse \(\theta(a,\kappa)\) can be up to \(\tfrac{\pi}{2}\), thus always sending the state back to \(\ket{\psi_0}\), preventing a gradual evolution of the angle. In such a case, the algorithm would lose its quantum speed-up. 
Interestingly, Corollary~\ref{cor:theta_leq_kappa} shows we can keep \(\theta(a,\kappa)\) small by bounding \(\kappa\), so that the action of \(G\) overcomes the effect of the collapse.

\begin{lemma} \label{lem:G_robustness}
The unitary map \(G\) is \(\varepsilon\)-robust to \(\kappa\)-measurements of \(\chi\) for
\begin{align}
  \kappa &\leq \tfrac{1}{\sqrt{\card{B}}} \\
  \varepsilon &= \sin 3\alpha \\
\intertext{which is witnessed by:}
  g(n) & = 2n
\end{align}
\end{lemma} 
\begin{proof}
Let \(\{a_n\}\) be the following sequence of angles:
\begin{equation}
\begin{aligned}
  a_{n+1} &= a_n + 2\alpha \\
  a_0     &= \alpha
\end{aligned}
\end{equation}
Then \(G^n\ket{\psi} = \ket{a_n}\), where \(\ket{\psi}\) is the initial state. Similarly, we define a sequence \(\{b_n\}\) satisfying \((\cW G)^n\ket{\psi} = \ket{b_n}\):
\begin{equation}\label{eq:recurrence}
\begin{aligned}
  b_{n+1} &= b_n - \theta(b_n,\kappa) + 2\alpha \\
  b_0     &= \alpha.
\end{aligned}
\end{equation}
Remember that \(\sin \alpha = \card{B}^{-1/2}\), so the imposed bound on \(\kappa\) can be rephrased as \(\kappa \leq \sin \alpha\). 
Thanks to Corollary~\ref{cor:theta_leq_kappa}, this implies \(\abs{\theta(b_n,\kappa)} \leq \alpha\) on any iteration.
Then \(\alpha \leq b_{n+1} - b_n \leq 3\alpha\) for any \(n \in \Nat\), whereas \(a_{n+1} - a_n = 2\alpha\).
Therefore
\begin{equation}
  \exists m \leq 2n\colon \quad b_m \leq a_n \leq b_{m+1}
\end{equation}
and \(a_n - b_m \leq 3\alpha\).
Because \(p_\chi(G^n\ket{\psi}) = \sin^2 a_n\) and \(p_\chi((\cW G)^m\ket{\psi}) = \sin^2 b_m\), it follows that
\begin{equation}
	\abs{p_\chi(G^n\ket{\psi}) - p_\chi((\cW G)^m\ket{\psi})} \ \leq \ \abs{\sin^2 a_n - \sin^2 (a_n \pm 3\alpha)}.
\end{equation}
Using \(\sin^2 x - \sin^2(x+y) = - \sin(y) \sin(2 x + y)\) we reach the conclusion that, for any \(n \in \Nat\)
\begin{equation}
  \exists m \leq 2n\colon \quad \abs{p_\chi(G^n\ket{\psi}) - p_\chi((\cW G)^m\ket{\psi})} \ \leq \ \abs{\sin 3\alpha}
\end{equation}
Hence, $G$ is \(\varepsilon\)-robust to \(\kappa\)-measurements for \(\varepsilon = \sin 3\alpha\), with witness \(g(n) = 2n\).
\end{proof}

From the general result of Lemma~\ref{lem:comp_guarantee} and the discussion following it, together with the assumption that \(\alpha\) is small (\(\sin 3\alpha \ll 1/2\)), we conclude that the algorithm in Figure~\ref{fig:code_walk} will succeed in finding the marked element \(\ket{\star}\) within
\begin{equation}
  T = g\left(f\left(\frac{2c}{\kappa(1-2\varepsilon)}\right)\right)
\end{equation}
iterations, with arbitrarily high probability, for small \(c \in \Nat\).
Combining this with Lemmas~\ref{lem:G_guarantee} and~\ref{lem:G_robustness} we find that, for \(\kappa \leq \card{B}^{-1/2}\), the number of iterations is within \(\bigO{\sqrt{\card{B}}}\). 

\begin{figure}
\centering
\includegraphics[scale=0.43]{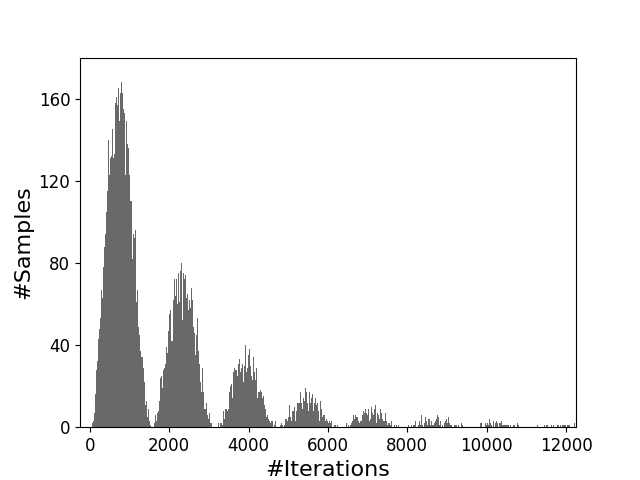}
\includegraphics[scale=0.43]{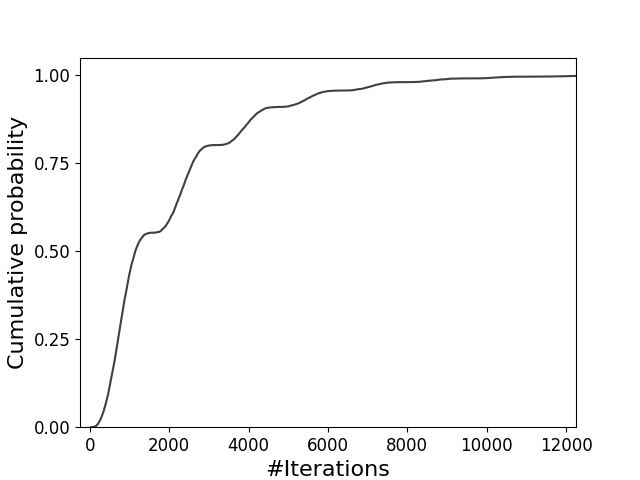}
\caption{Distribution of the number of iterations before success for parameters \(\card{B} = 10^6\) and \(\kappa = \card{B}^{-1/2}\); \emph{left:} histogram, \emph{right:} cumulative probability distribution. Drawn from \(10000\) samples, obtained by sampling the success probability \(p_\top = \kappa\sin^2{b_n}\) where \(b_n\) is given by~\eqref{eq:recurrence}. The median is approximately \(1000\) iterations, and mean \(2000\) iterations, \ie\ approximately \(\tfrac{1}{\kappa}\) and \(\tfrac{2}{\kappa}\) respectively.}
\label{fig:sampling}
\end{figure}

\begin{remark} \label{rmk:constant_factor}
Our analysis yields an arguably large constant factor
\begin{equation}
  T \approx (8c + \tfrac{\pi}{2})\sqrt{\card{B}}.
\end{equation}
However, it is important to remark that this factor is an overestimation, due to the simplifications applied when finding \(f\) and \(g\) in this section. With these simplifications we intended to prioritise clarity of our proof stategy. Tighter bounds would yield a more accurate factor. In fact, Figure~\ref{fig:sampling} suggests the average number of iterations is approximately \(2\sqrt{\card{B}}\).

Moreover, our choice of \(\kappa = \card{B}^{-1/2}\) is not optimal.
To prove Lemma~\ref{lem:G_robustness} we only needed that the collapse on each iteration was smaller than the increment of the angle due to \(G\), \ie\ for every angle \(a\), \(\theta(a,\kappa) < 2\alpha\). If we set \(\kappa = 5\card{B}^{-1/2}\), and use the bound of \(\theta(a,\kappa)\) given by Lemma~\ref{lem:theta_bound}, we can verify that, for this instance of \(\card{B} = 10^6\), \(\theta(a,\kappa) < 1.3\alpha\), so the collapse is appropriately bounded.
In this particular case, numerical analysis shows that the average number of iterations required is slightly smaller than \(\tfrac{\pi}{4}\sqrt{\card{B}}\), which is the number of iterations the standard Grover's algorithm runs for.
The optimal value of \(\kappa\) depends on the parameters of the problem (\ie in this case, it may be different for different values of \(\card{B}\)).
We conjecture that such an optimal value of \(\kappa\) exists for any instance of Grover's problem, so that our approach and the standard one coincide in their expected number of iterations.

The realisation that for a fine-tuned value of \(\kappa\) our algorithm would perform as well as the standard Grover's algorithm was first discussed in a recent paper~\cite{Differentiable}. The paper provides an automatic procedure, based on differentiable programming, to choose an optimal value of the parameters of a quantum program. It discusses our algorithm as an application of its approach and, for small values of \(\card{B}\), finds an optimal \(\kappa\) satisfying the conjecture above.
\end{remark}

Figure~\ref{fig:sampling} helps us visualise the behaviour of the algorithm: as the angle \(b_n\) monotonically increases throughout the iterations, the instantaneous probability of success \(p_\top = \kappa\sin^2{b_n}\) changes periodically. The peaks of the histogram correspond to intervals of \(\cA\)-iterations where the probability of success approaches its maximum (that is, \(p_\top \approx \kappa\)); these alternate with troughs where the probability approaches \(0\).

\section{Discussion}
\label{sec:discussion}

We have used the term \emph{active iteration} to refer to the stages of the loop where halting is most likely.
If the distribution of active iterations is not known, we cannot fix the number of iterations the loop should run for.
In such a situation, it is natural to use a while loop, but these require testing a termination condition, perturbing the state.
In this work, we have proposed \(\kappa\)-while loops, which let us keep the perturbation due to measurement below a threshold.
We have introduced the properties of \(\cA\)-guarantee and robustness, which let us identify the time complexity of algorithms using these \(\kappa\)-while loops.

We must emphasise that these \(\kappa\)-while loops can be realised in any quantum programming language with classical control flow, as indicated in Figure~\ref{fig:kappa_while}.
More precisely, \(\kappa\)-while loops are a particular instance of the while loops described in~\cite{YingLoop}.
Hence, we expect it would be immediate to apply to \(\kappa\)-while loops any findings from the literature on classical control flow of quantum programs; for instance, verification of program correctness~\cite{YingHoare}, study of loop termination~\cite{YingTermination,YingLoop} and program semantics~\cite{SelingerPL,YingBook}.
Considering the control flow is classical, there is no superposition of different branches of the computation exiting the loop at different stages.
However, the \(\kappa\)-while loop is not fully classical either, in the sense that some degree of superposition in the quantum data is maintained across iterations, which is enough to achieve quantum speed-up in certain situations.

The key component of the \(\kappa\)-while loop is a weak measurement parametrised by \(\kappa\). Weak measurements are not new to quantum computer scientists, as they are at the heart of quantum feedback control~\cite{QuantumFeedbackSurvey}. Our \(\kappa\)-while loop can be seen as an example of quantum feedback control where the weak measurements are applied at discrete time steps.
Discrete time feedback control has been used to protect a single qubit from decoherence~\cite{FeedbackQECC,FeedbackQECCExperiment}, while similar notions of weak measurement (over continuous time) have been proposed to monitor and drive complex evolutions~\cite{ContinuousFeedback}.
However, weak measurements are rarely used in the design of quantum algorithms; Mizel's work~\cite{Mizel} being the only case we are aware of.
Defining the \(\kappa\)-while loop programming construct introduces weak measurements to  algorithm designers and programming language experts in a language that is familiar to them.


\subsection{Comparison to other versions of Grover's algorithm}

We have shown that a \(\kappa\)-while loop may be used to implement Grover's algorithm; however, there is no algorithmic advantage with respect to the standard approach.
The reason is that the distribution of \(\cA\)-iterations throughout Grover's algorithm is easy to predict, and thus we can fix the number of iterations we should run it for in advance, instead of using a \(\kappa\)-while loop.
Our approach can be easily extended to the setting of amplitude amplification~\cite{AAA}.

In principle, our \(\kappa\)-while loop method (as well as Mizel's~\cite{Mizel}) requires knowledge of the proportion \(\gamma\) of marked states in the database (in Grover's, \(\gamma = 1/B\)) so that \(\kappa \approx \gamma^{-1/2}\) can be chosen appropriately.
However, the tricks used in the standard algorithm to figure out \(\gamma\) can also be applied to our setting.
Interestingly, whereas an under- or overestimation of \(\gamma\) causes the standard algorithm to behave unpredictably, our approach will always terminate in success.
In particular, if we underestimate the proportion and set \(\kappa \ll \gamma^{-1/2}\), the quadratic speed-up will be maintained, but we will run for more iterations than strictly necessary.
On the other hand, if we overestimate it and set \(\kappa \gg \gamma^{-1/2}\), there will be too much collapse per iteration, resulting in the loss of quantum speed-up.

The latter point was already argued by Mizel~\cite{Mizel}, who proposed a version of Grover's algorithm that is, in essence, the same as ours.
Mizel presents the algorithm as a fixed point routine, where the state gradually converges to a marked state throughout the iterations.
However, this fixed point behaviour is an artefact of disregarding the outcome of the weak measurement in their analysis (see Remark~\ref{rmk:known_probe_outcome}).
In reality, the state keeps evolving at the same rate throughout the algorithm until a \(\top\) measurement outcome occurs, when the state collapses onto the marked subspace.
The field of quantum trajectories~\cite{ToddTutorial} focuses on understanding these kind of stochastic dynamics in the presence of weak measurements.

It is worth mentioning that algorithms for Grover's search using a fixed point approach do exist~\cite{GroverFixPoint,YoderFixPoint}.
In these, no measurement is applied during execution; instead, each iteration applies a unitary parametrised by a value that is gradually reduced throughout the algorithm.
Intuitively, each iteration moves the state closer to a marked one, but each time the step is smaller to avoid overshooting.
The drawback of this fixed point approach is that, unless we can implement a circuit where the unitary's parameter can be changed during runtime, each iteration requires a different circuit.

\subsection{Applications beyond quantum search}
\label{sec:beyond_quantum_search}

We have presented a \(\kappa\)-while loop version of Grover's algorithm as a proof of concept.
The next step in this project is to find practical uses of \(\kappa\)-while loops: quantum algorithms where the distribution of active iterations is unknown or it would be costly to predict.
In Section~\ref{sec:motivation} we have suggested that choosing \(\kappa\)-while loops over for loops may make algorithms more reliable against unpredictable behaviour in the body of the loop.
It would be valuable to find an explicit example of a quantum algorithm that, under a realistic noise model, behaves more reliably when implemented using \(\kappa\)-while loops.

Another natural line of research is to identify practical problems where the distribution of \(\cA\)-iterations is inherently unpredictable.
The literature on quantum walks provides multiple results where quantum computers exhibit an advantage in the study of Markov processes~\cite{Szegedy,VertexFinding}.
The unitary evolution of these quantum walks is defined in terms of the transition matrix of the Markov process, which is often only required to be symmetric and ergodic.
Thus, the literature from this field provides us with a diverse set of examples of quantum algorithms whose evolution may be arbitrarily complex.
We argue there may be Markov processes whose quantum walk can be proven to be robust to \(\kappa\)-measurements as in Definition~\ref{def:robustness} while, at the same time, too unpredictable to let us estimate when the evolution reaches a desired state.
The argument goes as follows: on one hand, the choice of \(\kappa\) will be determined by a lower bound of the rate at which the walker traverses the graph (so that \(\kappa\)-measurements are not so strong that they would prevent the walker from reaching certain regions of the graph); this rate may be estimated as the infimum of local rates, using notions such as conductance and effective resistances~\cite{Belovs}.
On the other hand, in order to predict when the walker reaches a particular state, we need to take into account the global behaviour of the walk, which may be a more daunting task.
We therefore find the field of~\cite{Szegedy,VertexFinding,Belovs} to be a particularly promising area where to look for applications of \(\kappa\)-while loops.

The notion of weak measurement used in our definition of \(\kappa\)-while loop is discrete. 
However, continuous measurements have been studied extensively~\cite{ContinuousMeas} and the body of the while loop itself may be replaced by a continuous evolution given by a Hamiltonian, thus extending \(\kappa\)-while loops to the continuous-time setting.
This may be valuable from an experimentalist perspective; for instance, it hints at implementations of algorithms (for instance, Grover's search) where the probe is continuously measured throughout the execution, thus simplifying control, \eg\ we do not need to know when each iteration finishes and the next one starts.

The \(\kappa\)-while loop is a programming construct that offers a promising abstraction: a way to iteratively `peek' at quantum states without completely collapsing them. The key challenge to its use in algorithms is the need to strike a balance -- by tuning parameter \(\kappa\) -- so that the collapse is low enough, while the information gained is sufficient. The body of a \(\kappa\)-while loop may be any CP map, hence \(\kappa\)-while loops may be nested inside each other. As with classical while loops, termination implies the satisfaction of the predicate, which is a useful feature for the analysis of program correctness.

\paragraph{Acknowledgements.} Pablo Andres-Martinez is supported by EPSRC grant EP/L01503X/1 via the Centre for Doctoral Training in Pervasive Parallelism at the University of Edinburgh, School of Informatics. Chris Heunen is supported by EPSRC Fellowship EP/R044759/1.

\bibliographystyle{plainnat}
\bibliography{Bibliography}

\appendix

\section{A bound to the collapse}
\label{app:bound_proof}

\begin{lemma*}
For any \(\kappa\) and angle \(a\)
\begin{equation} \notag
  \abs{\theta(a,\kappa)} \leq \arcsin \left( \frac{1-\xi}{1+\xi} \right).
\end{equation}
\end{lemma*}
\begin{proof}

Suppose \(a \in [0,\tfrac{\pi}{2}]\) and use the properties of sines on the triangle from Figure~\ref{fig:theta_triangle}i to obtain:
\begin{equation}
  \frac{\sin (a - \theta)}{\xi \sin a} = \frac{\sin (\pi/2 - a + \theta)}{\cos a}
\end{equation}
This simplifies as follows:
\begin{align}
 \begin{aligned}
  && \frac{\sin (a - \theta)}{\xi \sin a} &= \frac{\cos (a - \theta)}{\cos a} \\
  \iff&& \tan (a - \theta) &= \xi \tan a \\
  \iff&& \frac{\tan a - \tan \theta}{1 + \tan a \tan \theta} &= \xi \tan a \\
 \end{aligned}
\end{align}
Solving for \(\theta\) gives:
\begin{equation}\label{eq:theta}
  \theta(a,\kappa) = \arctan \left( \frac{(1 - \xi) \tan a}{1 + \xi \tan^2 a} \right)
\end{equation}

If \(a \in \left[\tfrac{\pi}{2},\pi\right]\) instead, a similar analysis yields:
\begin{equation}\label{eq:thetaNeg}
  \theta(a,\kappa) = -\arctan \left( \frac{(1 - \xi) \tan a}{1 + \xi \tan^2 a} \right)
\end{equation}
which only differs from equation~\eqref{eq:theta} in the sign. If \(a\) is in the third quadrant, then we obtain equation~\eqref{eq:theta} and, if \(a\) is in the fourth quadrant, we get~\eqref{eq:thetaNeg}.
These sign changes are convenient: the geometric analysis in Figure~\ref{fig:theta_triangle} shows that in the first (and third) quadrant, the resulting angle is \(a - \abs{\theta}\), whereas in the second (and fourth) quadrant it is \(a + \abs{\theta}\). Thus the resulting angle is \(a - \theta\) regardless of which quadrant \(a\) lies in.

Next, we determine the maximum value of \(\theta(a,\kappa)\). To do so, fix \(\kappa\) and find the critical points of \(\theta\) as a function on \(a\). These happen where the derivative
\begin{equation}
  \frac{\mathrm{d}\theta}{\mathrm{d}a} = 1 - \frac{\xi}{\cos^2 a + \xi^2\sin^2 a}
\end{equation}
vanishes. Critical points occur periodically (once in every quadrant), but all of them reach the same absolute value. The critical point within the first quadrant happens at:
\begin{equation}\label{eq:max_at_a}
  a = \arccos \left( \sqrt{\frac{\xi}{\xi + 1}} \right)
\end{equation}
Applied to equation~\eqref{eq:theta} this gives a tight upper bound:
\begin{equation}\label{eq:theta_bound_ugly}
  \abs{\theta(a,\kappa)} \leq \arctan \left( \frac{1-\xi}{2\sqrt{\xi}} \right)
\end{equation}
Using the equality \(\sin(\arctan(x)) = \frac{x}{\sqrt{x^2 + 1}}\), the fact that \(\xi \in [0,1]\) and some basic algebra we reach the claim:
\begin{equation}
  \sin \abs{\theta(a,\kappa)} \leq \frac{1-\xi}{1+\xi}.
\end{equation}

\end{proof}

\end{document}